\documentclass{iopart}
\usepackage{verbatim}

\begin{document}

\title[]{Symmetric hyperbolic Euler equations for relativistic perfect fluids}
\author[]{Ronald A. Walton}
\address {1671 Via Rancho, San Lorenzo, CA 94580, rwsailor@mac.com} 



 \begin{abstract}
	The Euler equations governing a relativistic perfect fluid are put into symmetric hyperbolic form with dependent variables the fluid's specific entropy plus a generalized velocity vector equal to the fluid's unit relativistic velocity vector divided pointwise by the fluid's specific enthalpy.
\end{abstract}

	\noindent \textbf{Introduction}.  Frauendiener \cite{JF03} and Walton \cite{RAW03} independently derived covariant symmetric hyperbolic Euler equations for isentropic relativistic perfect fluids.   The purpose of this paper is to derive covariant symmetric hyperbolic Euler equations for the more general, nonisentropic relativistic perfect fluids.
	
	The dependent variables of a relativistic perfect fluid include a unit velocity vector field $U^{\alpha}$ plus two thermodynamic scalar fields which I will choose to be the fluid's specific enthalpy $h$ and specific entropy $s$.  Following Frauendiener and Walton, $U^{\alpha}$ and $h$ are merged into a generalized velocity vector field $J^{\alpha} = U^{\alpha} / h$.  Then, the Euler equations for the new dependent variables $J^{\alpha}$ and $s$ are derived from the assumptions that energy-momentum and material are locally conserved by the fluid's motions.  These Euler equations are not symmetric hyperbolic, but they imply the fluid's specific entropy $s$ is necessarily a first integral of $J^{\alpha}$.  The latter is used to rewrite the Euler equations in an equivalent symmetric hyperbolic form.		
	Having symmetric hyperbolic Euler equations for a relativistic perfect fluid is important for two reasons.  First, existence and uniqueness of solutions to the Euler equations are immediately guaranteed by symmetric hyperbolicity.  Indeed, proving existence and uniqueness for solutions of the usual non-symmetric hyperbolic Euler equations is difficult;  see Lichnerowicz \cite{AL67} for examples.  And, second, to numerically solve the Euler equations, symmetric hyperbolic equations are preferred.
	
	In this paper Greek letters are used to denote tensor indices and units are used in which the \emph{speed of light} $ = 1$.
	
\medskip	
	\noindent \textbf{Thermodynamics}.  In this paper always assume a relativistic perfect fluid obeys an equation of state $p = p(h,s)$, defining the fluid's pressure $p$ to be a function of the fluid's specific enthalpy $h$ and specific entropy $s$.  The  fundamental relation of thermodynamics,
\begin{equation} \label{frt}
dp = n \; dh - n T \; ds,
\end{equation}
then informs us that the fluid's material density $n$ and absolute temperature $T$ can be derived from the equation of state by the equations:
\begin{equation} \label{n}
n = \frac{\partial p}{\partial h}
\end{equation}
and
\begin{equation} \label{T}
T = - \frac{1}{n} \frac{\partial p}{\partial s}.
\end{equation}
And, from the definition of the fluid's energy density $e = n h - p$, the scalar field $v$ derived by the equation:
\begin{equation} \label{sos}
v^{-2} = \left(\frac{\partial e}{\partial p}\right)_s = \frac{\partial e/\partial h}{\partial p/\partial h} = \frac{h}{n} \frac{\partial n}{\partial h} = \frac{h}{n} \frac{\partial^2 p}{\partial h^2},
\end{equation}
is the speed of sound in the fluid.
\medskip

	\noindent \textbf{Constitutive Functions}.  Let $g_{\alpha \beta}$ be the spacetime's metric tensor and assume it has signature $-+++$.  And, following Walton \cite{RAW03} and Frauendiener \cite{JF03}, let $J^{\alpha}$ be a vector field tangent to the timelike fluid flow and normalized to define the value of the fluid's specific enthalpy by the equation
\begin{equation} \label{h}
h = (- g_{\alpha \beta} J^{\alpha} J^{\beta})^{-1/2}.
\end{equation}
Then the relativistic velocity vector field $U^{\alpha}$, energy-momentum tensor field $T_{\alpha}^{\beta}$, and material current vector field $N^{\alpha}$ of a relativistic perfect fluid are the following functions of the fields $J^{\alpha}$ and $s$:
\begin{equation} \label{U}
U^{\alpha} = h J^{\alpha},
\end{equation}
\begin{equation} \label{emt}
T_{\alpha}^{\beta} = n h^3 J_{\alpha} J^{\beta} + p g_{\alpha}^{\beta},
\end{equation}
\begin{equation} \label{mcv}
N^{\alpha} = n h J^{\alpha}.
\end{equation}
And the tensor field
\begin{equation} \label{q}
q_{\alpha \beta} = g_{\alpha \beta} + h^2 J_{\alpha} J_{\beta}
\end{equation}
has the useful algebraic property of projecting vector fields onto their component orthogonal to $J^{\alpha}$:  $q^{\alpha}_{\beta} J^{\beta} = 0$.
\medskip 

	\noindent \textbf{Some Calculus}.  Let $\nabla_{\mu}$ denote the covariant derivative operator with symmetric connection defined by the Christoffel symbols of the metric tensor $g_{\alpha \beta}$, so $\nabla_{\mu} g_{\alpha \beta} = 0$.  Using equations (\ref{frt}), (\ref{n}), (\ref{sos}), and (\ref{h}) we can derive:
\begin{equation} \label{dh}
\nabla_{\mu} h = h^3 J_{\beta} \nabla_{\mu} J^{\beta},
\end{equation}
\begin{equation} \label{dn}
\nabla_{\mu} n = \frac{\partial n}{\partial h} \nabla_{\mu} h + \frac{\partial n}{\partial s} \nabla_{\mu} s = nh^2 v^{-2} J_{\beta} \nabla_{\mu} J^{\beta} + \frac{\partial n}{\partial s} \nabla_{\mu} s,
\end{equation}
\begin{equation} \label{dp}
\nabla_{\mu} p = n \nabla_{\mu} h - n T \nabla_{\mu} s = n h^3 g_{\mu}^{\nu} J_{\beta} \nabla_{\nu} J^{\beta} - n T \nabla_{\mu} s.
\end{equation}
And, then from the definitions of the energy-momentum tensor field and the material current vector field we can compute the covariant divergences:
\begin{eqnarray} \label{demt}
\nabla_{\mu} T_{\alpha}^{\mu} &=& \nabla_{\mu} (n h^3 J_{\alpha} J^{\mu} + p g_{\alpha}^{\mu}) \nonumber \\
&=& (\nabla_{\mu} n) h^3 J_{\alpha} J^{\mu} + 3 n h^2 (\nabla_{\mu} h) J_{\alpha} J^{\mu} \nonumber \\
&&+ n h^3 (g_{\alpha \beta} J^{\mu}  + J_{\alpha} g_{\beta}^{\mu}) \nabla_{\mu} J^{\beta} + n g_{\alpha}^{\mu} \nabla_{\mu} h - n T g_{\alpha}^{\mu} \nabla_{\mu} s \nonumber \\
&=&nh^3( v^{-2} h^2 J_{\alpha} J_{\beta} J^{\mu} + 3 h^2 J_{\alpha} J_{\beta} J^{\mu} 
+ g_{\alpha \beta} J^{\mu}  \nonumber\\
&&+ J_{\alpha} g_{\beta}^{\mu} + g_{\alpha}^{\mu} J_{\beta} ) \nabla_{\mu} J^{\beta} + \left( \frac{\partial n}{\partial s} h^3 J_{\alpha} J^{\mu} - n T g_{\alpha}^{\mu} \right) \nabla_{\mu} s \nonumber \\
&=&nh^3( v^{-2} h^2 J_{\alpha} J_{\beta} J^{\mu} + q_{\alpha \beta} J^{\mu}  + J_{\alpha} q_{\beta}^{\mu} + q_{\alpha}^{\mu} J_{\beta} ) \nabla_{\mu} J^{\beta} \nonumber \\
& & + \left( \frac{\partial n}{\partial s} h^3 J_{\alpha} J^{\mu} - n T g_{\alpha}^{\mu} \right) \nabla_{\mu} s, 
\end{eqnarray}
and:
\begin{eqnarray} \label{dmcv}
\nabla_{\mu} N^{\mu} &=& \nabla_{\mu} (n h J^{\mu})  \nonumber \\
&=& (\nabla_{\mu} n) h J^{\mu} + n (\nabla_{\mu} h) J^{\mu} + n h g_{\beta}^{\mu} \nabla_{\mu} J^{\beta} \nonumber  \\
&=& n h ( v^{-2} h^2 J_{\beta} J^{\mu} + h^2 J_{\beta} J^{\mu} + g_{\beta}^{\mu} ) \nabla_{\mu} J^{\beta} + \frac{\partial n}{\partial s} h J^{\mu} \nabla_{\mu} s  \nonumber \\
&=& n h ( v^{-2} h^2 J_{\beta} J^{\mu} + q_{\beta}^{\mu} ) \nabla_{\mu} J^{\beta} + \frac{\partial n}{\partial s} h J^{\mu} \nabla_{\mu} s.  
\end{eqnarray}

	\noindent \textbf{Euler Equations}.  The Euler equations are derived from the assumption that the fluid's energy-momentum tensor field and material current vector field both have zero covariant divergence, so energy-momentum and material are each locally conserved.  From equations (\ref{demt}) and (\ref{dmcv}), the Euler equations are explicitly:
\begin{eqnarray} \label{lemc}
n h^3 (v^{-2} h^2  J_{\alpha} J_{\beta} J^{\mu} &+& q_{\alpha \beta} J^{\mu} + J_{\alpha} q_{\beta}^{\mu} + q_{\alpha}^{\mu} J_{\beta} ) \nabla_{\mu} J^{\beta} \nonumber \\
&&+ \left( \frac{\partial n}{\partial s} h^{3} J_{\alpha} J^{\mu}  - nT g_{\alpha}^{\mu} \right) \nabla_{\mu} s = 0,
\end{eqnarray}
and: 
\begin{equation} \label{lmc}
n h (v^{-2} h^2   J_{\beta} J^{\mu} + q_{\beta}^{\mu} ) \nabla_{\mu} J^{\beta}  + \frac{\partial n}{\partial s}  h J^{\mu} \nabla_{\mu} s = 0.
\end{equation}
Equations (\ref{lemc}) and (\ref{lmc}) are not symmetric, and therefore not symmetric hyperbolic, because the coefficient of $\nabla_{\mu} s$ in (\ref{lemc}) and the coefficient of $\nabla_{\mu} J^{\beta}$ in (\ref{lmc}) are different.\medskip

	\noindent \textbf{Symmetric Euler Equations}.  From the Euler equations (\ref{lemc}) and (\ref{lmc}) it follows that the specific entropy $s$ is a first integral of the vector field $J^{\alpha}$:
\begin{equation} \label{lec}
J^{\alpha} \nabla_{\mu} T_{\alpha}^{\mu} + \nabla_{\mu} (nh J^{\mu}) = - n T J^{\mu} \nabla_{\mu} s = 0,
\end{equation}
so a relativistic perfect fluid flows adiabatically.  Multiples of $J^{\mu} \nabla_{\mu} s$ can therefore be added to/subtracted from the Euler equations (\ref{lemc}) and (\ref{lmc}) and the new equations will be \emph{equivalent} to the Euler equations, if the new equations also imply $s$ is a first integral of $J^{\alpha}$.\footnote{Equivalent means fields $J^{\alpha}$ and $s$ which solve the new equations will also solve the Euler equations (\ref{lemc}) and (\ref{lmc}), and conversely.}  In particular, subtracting $[(\partial n/\partial s) h + (1 + v^{-2}) n T] h^2 J_{\alpha} J^{\mu} \nabla_{\mu} s$  and $[(\partial n/\partial s)h + (1 + v^{-2}) n T ] J^{\mu} \nabla_{\mu} s$ from equations (\ref{lemc}) and (\ref{lmc}), respectively, and multiplying the resulting second equation by $-T/h$ will symmetrize the Euler equations.  Equations (\ref{lemc}) become:
\begin{eqnarray} \label{sh1}
n h^3 ( v^{-2} h^2 J_{\alpha} J_{\beta} J^{\mu}  + q_{\alpha \beta} J^{\mu} &+& J_{\alpha} q_{\beta}^{\mu} + q_{\alpha}^{\mu} J_{\beta} ) \nabla_{\mu} J^{\beta} \nonumber \\ 
&-& n T ( v^{-2} h^2 J_{\alpha} J^{\mu}  + q_{\alpha}^{\mu}) \nabla_{\mu} s = 0;
\end{eqnarray}
and equation (\ref{lmc}) becomes:
\begin{equation} \label{sh2}
- n T (v^{-2} h^2 J_{\beta} J^{\mu}  + q_{\beta}^{\mu}) \nabla_{\mu} J^{\beta} + n T^2 h^{-1} (1 + v^{-2}) J^{\mu} \nabla_{\mu} s = 0.
\end{equation}
Because the linear combination of equations (\ref{sh1}) and (\ref{sh2}):
\[
J^{\alpha}\cdot(\ref{sh1})_{\alpha} - h T^{-1}\cdot(\ref{sh2}) = - n T  J^{\mu} \nabla_{\mu} s = 0,
\]
implies the specific entropy is a first integral of the vector field $J^{\alpha}$, the symmetric equations  (\ref{sh1}) and (\ref{sh2}) are equivalent to the Euler equations (\ref{lemc}) and (\ref{lmc}).  Equations (\ref{sh1}) and (\ref{sh2}) will be referred to as the \emph{symmetric Euler equations}.
\medskip

	\noindent \textbf{Isentropic Euler equations}.  A relativistic perfect fluid is isentropic if its specific entropy $s = constant$.  According to equation (\ref{lec}), in an isentropic perfect fluid local material conservation (\ref{lmc})  is a consequence of local energy-momentum conservation (\ref{lemc}).  An isentropic relativistic perfect is therefore governed by the reduced system of equations:
\begin{equation} \label{ilemc}
n h^3 ( v^{-2} h^2 J_{\alpha} J_{\beta} J^{\mu}  + q_{\alpha \beta} J^{\mu} + J_{\alpha} q_{\beta}^{\mu} + q_{\alpha}^{\mu} J_{\beta} ) \nabla_{\mu} J^{\beta} = 0,
\end{equation}
with only the field $J^{\alpha}$ as a dependent variable.  Frauendiener \cite{JF03} and Walton \cite{RAW03} independently derived equations (\ref{ilemc}) and proved they are symmetric hyperbolic.
\medskip

	\noindent \textbf{Characteristic Analysis}.  Introduce an orthonormal vector field basis consisting of the unit fluid velocity vector field $U^{\alpha} = h J^{\alpha}$ plus three unit spacelike vector fields $X^{\alpha}$, $Y^{\alpha}$, and $Z^{\alpha}$.  And let $\xi^{\mu} = \omega U^{\mu} + \kappa^{\mu}$, $\kappa^{\mu} U_{\mu} = 0$, be an arbitrary vector field.  There is no loss of generality by assuming the vector field $\kappa^{\mu} =  \kappa X^{\mu}$, $\kappa > 0$.  Then, with respect to this basis and to the vector field $\xi^{\mu}$, the characteristic matrix for the symmetric Euler equations (\ref{sh1}) and (\ref{sh2}) is the symmetric $5 \times 5$ matrix:
\begin{equation} \label{cm}
\mathbf{A}_{\xi} \!=\! - n h^2 \! \! \left(\begin{array}{ccccc}
v^{-2} \omega & \kappa & 0 & 0 & T h^{-2}  v^{-2} \omega \\
 \kappa & \omega & 0 & 0 & T h^{-2} \kappa \\
0 & 0&  \omega & 0 & 0 \\
0 & 0 & 0 &  \omega & 0 \\
T h^{-2} v^{-2} \omega & T h^{-2} \kappa & 0 & 0 & T^2 h^{-4} (1 \!+\! v^{-2}) \omega
\end{array} \right)\!\!. 
\end{equation}
And the characteristic polynomial equation (dispersion relation) for equations (\ref{sh1}) and (\ref{sh2}) is:
\begin{equation} \label{dr}
\det{(\mathbf{A}_{\xi})} = - n^5 h^{6} T^2  \omega^3 (v^{-2} \omega^2 - \kappa^2) = 0,
\end{equation}
where $\det(-)$ denotes the determinant operator for square matrices.  Obviously, for each \emph{wave vector} $\kappa^{\mu} = \kappa X^{\mu}$ the characteristic polynomial equation has only real roots in the \emph{frequency} variable $\omega$, if the speed of sound $v $ is real valued:  $\omega_1(\kappa) = \omega_2(\kappa) = \omega_3(\kappa) = 0$,  $\omega_4(\kappa) = v \kappa = v \sqrt{\kappa^{\mu} \kappa_{\mu}}$, and $\omega_5(\kappa) = - v \kappa = - v \sqrt{\kappa^{\mu} \kappa_{\mu}}$.  Therefore, according to the theory of partial differential equations (Courant \& Hilbert \cite{CH62}):   \emph{equations (\ref{sh1}) and (\ref{sh2}) are hyperbolic with wave modes propagating in the bicharacteristic directions}:
\begin{equation}
\chi^{\alpha}_I = U^{\alpha} + g^{\alpha \mu} \frac{\partial \omega_I(\kappa)}{\partial \kappa^{\mu}} = U^{\alpha} + \frac{\omega_I(\kappa)}{\kappa} X^{\alpha}, \;  I = 1,2,3,4,5,
\end{equation}
\emph{tangent to three dimensional characteristic hypersurfaces with normal vector fields}:
\begin{equation}
\xi_I^{\mu} = \omega_I(\kappa) U^{\mu} + \kappa X^{\mu}, \;  I = 1,2,3,4,5.
\end{equation}
Such wave modes propagate at \emph{speeds} $V_I(\kappa) = |\omega_I(\kappa)| / \kappa$, $I=1,2,3,4,5$, with respect to observers comoving with the fluid at the relativistic velocity $U^{\alpha}$.  The modes $I=1,2,3$ are stationary with respect to the fluid;  they represent flow along the fluid's streamlines.   The modes $I=4,5$ propagate at the speed $v$ with respect to the fluid;  they represent sound waves, because no other wave modes are possible in a perfect fluid, and justify interpreting $v$ as the speed of sound in the fluid (Taub \cite{AT48}, Lichnerowicz \cite{AL67}.).
\medskip

	\noindent \textbf{Symmetric Hyperbolicity}.  As already noted, the characteristic matrix (\ref{cm}) of equations (\ref{sh1}) and (\ref{sh2}) is symmetric.  Let $\mathbf{W}$ be a real valued, $5 \times 1$ matrix with transpose $\mathbf{W}^T = (W_1,W_2,W_3,W_4,W_5)$.  Then equations (\ref{sh1}) and (\ref{sh2}) will be symmetric hyperbolic with respect to a vector field $\xi^{\mu}$ if the quadratic form defined by the characteristic matrix (\ref{cm}):
\begin{eqnarray} \label{qf}
\; \; \mathbf{W}^T \cdot \mathbf{A}_{\xi} \cdot \mathbf{W} = -nh^2 \{ [&v&^{-2}(W_1 + Th^{-2}W_5)^2 + W_2^2 + W_3^2 + W_4^2 \nonumber \\
&&+ T^2h^{-4} W_5^2] \omega \nonumber \\
&& + 2 (W_1 + Th^{-2}W_5) W_2 \kappa  \},
\end{eqnarray}
is definite for all $\mathbf{W} \ne 0$ (Courant \& Hilbert \cite{CH62}, Friedrichs \cite{KOF74}).

	Suppose the vector field $\xi^{\mu} = \omega U^{\mu} + \kappa X^{\mu}$ satisfies the inequality $(v^2 q_{\mu \nu}  - h^2 J_{\mu} J_{\nu}) \xi^{\mu} \xi^{\nu} = v^2 \kappa^2 - \omega^2 < 0$.  The latter is equivalent to $|\omega| >  v \kappa \ge0$, implying $\omega \ne 0$.   Definiteness of the quadratic form (\ref{qf}) then follows because: 
\begin{eqnarray*}
- \mathbf{W}^T \cdot \mathbf{A}_{\xi} \cdot \mathbf{W} / (\omega n h^2) &=& v^{-2}(W_1 +Th^{-2}W_5)^2 + W_2^2 + W_3^2 + W_4^2 \\
&&+ T^2h^{-4} W_5^2 + 2 (W_1 + Th^{-2}W_5) W_2 \kappa / \omega \\
&=& ( |W_1 + Th^{-2}W_5|/v - |W_2| )^2 \\
&& + 2 |W_1 + Th^{-2}W_5| |W_2| /v + W_3^2 + W_4^2 \\
&&+  T^2h^{-4} W_5^2 + 2 (W_1 + Th^{-2}W_5) W_2 \kappa / \omega,
\end{eqnarray*}
has a lower bound which is a sum of nonnegative summands with at least one summand always nonzero if $\mathbf{W} \ne 0$:
\begin{eqnarray*}
- \mathbf{W}^T \cdot \mathbf{A}_{\xi} \cdot \mathbf{W} / (\omega n h^2) &\ge& ( |W_1 + Th^{-2}W_5|/v - |W_2| )^2 + W_3^2 + W_4^2 \\
&&+  T^2h^{-4} W_5^2 \\
&&+ 2 |W_1 + Th^{-2}W_5| |W_2| (1/v - \kappa/|\omega|).
\end{eqnarray*}  
Here $|\cdot|$ denotes the absolute value of the enclosed quantity.  

	Therefore:  \emph{if the speed of sound $v$ in a relativistic perfect fluid is real valued, then the symmetric Euler equations (\ref{sh1}) and (\ref{sh2}) are symmetric hyperbolic with respect to any vector field $\xi^{\mu}$ which satisfies the inequality:  $(v^2 q_{\mu \nu}  - h^2 J_{\mu} J_{\nu}) \xi^{\mu} \xi^{\nu} < 0$.}
	
	Interestingly, symmetric hyperbolicity of the Euler equations puts no limit on the speed of sound in a relativistic perfect fluid other than it be finite.  But physics does:  in any physically reasonable relativistic perfect fluid, the speed of sound should not exceed the speed of light:  $v \le 1$.
\medskip
	
	\noindent \textbf{The Cauchy Problem}.  For any scalar field $\tau$ such that $\nabla_{\mu} \tau \ne 0$, the equation $\tau = constant$ defines a local foliation of spacetime by three dimensional hypersurfaces with normal vector field $g^{\mu \nu} \nabla_{\nu} \tau$.  The Cauchy Problem for the symmetric Euler equations (\ref{sh1}) and (\ref{sh2}) asks:  does there exist a unique set of fields $J^{\alpha}$ and $s$  which both satisfy equations (\ref{sh1}) and (\ref{sh2}) and take freely specified values $\bar{J}^{\alpha} = J^{\alpha}|_{\tau = 0}$ and $\bar{s} = s|_{\tau = 0}$ on the initial value hypersurface  $\tau = 0$ of the foliation?   From the works of Friedrichs \cite{KOF54}, Courant \& Hilbert \cite{CH62}, Fischer \& Marsden \cite{FM72,FM79},  and others:  \emph{if equations (\ref{sh1}) and (\ref{sh2}) are symmetric hyperbolic with respect to the normal vector field $\xi^{\mu} = g^{\mu \nu} \nabla_{\nu} \tau$ of the foliation, and if the initial values $\bar{J}^{\alpha}$ and $\bar{s}$ have square integrable derivatives of order $N \ge 4$ on the initial value hypersurface, then the answer to the Cauchy Problem for equations (\ref{sh1}) and (\ref{sh2}) is yes}. \medskip
	
	\noindent \textbf{A Final Note}  The symmetric Euler equations (\ref{sh1}) and (\ref{sh2}) for the fields $J^{\alpha}$ and $s$ are not uniquely symmetric hyperbolic.  For any positive scalar field $\theta$, one can add $(\theta - 1) n T^2 h^{-1} J^{\mu} \nabla_{\mu} s$ to the left hand side of (\ref{sh2}) and obtain:
\begin{equation} \label{sh2a}
- n T (v^{-2} h^2 J_{\beta} J^{\mu}  + q_{\beta}^{\mu}) \nabla_{\mu} J^{\beta} + n T^2 h^{-1} (\theta + v^{-2}) J^{\mu} \nabla_{\mu} s = 0.
\end{equation}
The system of equations (\ref{sh1}) and (\ref{sh2a}) is symmetric, equivalent to the Euler equations (\ref{lemc}) and (\ref{lmc}), and symmetric hyperbolic with respect to any vector field $\xi^{\mu}$ which satisfies the inequality:  $(v^2 q_{\mu \nu}  - h^2 J_{\mu} J_{\nu}) \xi^{\mu} \xi^{\nu} < 0$.  Fields $J^{\alpha}$ and $s$ which satisfy equations (\ref{sh1}) and (\ref{sh2a}) therefore also satisfy the equations (\ref{lemc}) and (\ref{lmc}), or equations (\ref{sh1}) and (\ref{sh2}), and conversely, and so must be independent of choice for the arbitrary positive scalar field $\theta$.  Thus, there is no loss of generality by setting $\theta = 1$ and considering equations (\ref{sh1}) and (\ref{sh2}) to be the symmetric hyperbolic Euler equations for relativistic perfect fluids.

\end{document}